\documentclass[twocolumn,showpacs,preprintnumbers,amsmath,amssymb]{revtex4}
\usepackage{graphicx}
\usepackage{dcolumn}
\usepackage{bm}
\begin{document}
\title{Field enhancement factor and field emission from\\ a hemi-ellipsoidal metallic needle}
\author{Evgeny G. Pogorelov}
\author{Alexander I. Zhbanov}
\email{azhbanov@gate.sinica.edu.tw}
\author{Yia-Chung Chang}
\affiliation{Research Center for Applied Sciences, Academia Sinica, 128,
  Section 2, Academia Road Nankang, Taipei 115, Taiwan.}
\date{\today}

\begin{abstract}
We present an exact solution for the electrostatic field between a metallic hemi-ellipsoidal
needle on a plate (as a cathode) and a flat anode. The basic idea is to replace the cathode
by a linearly charged thread in a uniform electric field and to use a set of ``image'' charges
to reproduce the anode. We calculate the field enhancement factor on the needle surface and
ponderomotive force acting on the needle. Using the Fowler-Nordheim theory we obtain an exact
analytical formula for the total current.
\end{abstract}
\pacs{7970+g, 8105Tp}
\maketitle

\section{\label{sec:level1}Introduction}

Application of various one-dimensional nanostructure materials as field emission sources has attracted extensive scientific efforts. Elongated structures are suitable for achieving high field-emission-current density at a low electric field because of their high aspect ratio. Area of its application includes a wide range of field-emission-based devices such as flat-panel displays, electron microscopes, vacuum microwave amplifiers, X-ray tube sources, cathode-ray lamps, nanolithography systems, etc.

Since the discovery of carbon nanotubes (CNT) [1-3] and experimental observations of their remarkable field emission characteristics [4-6], significant efforts have been devoted to the application of using CNT for electron sources. Recently, field emission from metals [7], metal oxides [8-11], metal carbides [12], and other elongated nanostructures have also been explored. It is now possible to control the diameter, height, radius of curvature of the tip, and basic form of emitters during growth. Elongated structures of different shapes such as nanotubes, nanocones, nanofibers, nanowires, nanoneedles, and nanorods have been successfully grown [13-17].

The current emitted by elongated nanostructures depends on the distribution of field strength at its top. A great deal of attention of researchers has been devoted to the apex field enhancement factor. Various approximations were used to evaluate the field amplification. Typically, the field enhancement factor is calculated from the slope of the Fowler–Nordheim plot. Under such consideration one can figure out only the field enhancement factor averaged over the entire emitting surface. Various theoretical approaches based on both analytical solutions and numerical simulations have been reported.[18-24] Analytical solution was obtained for a hemisphere on a plane, a floating sphere at the emitter-plane; a hemi-ellipsoid on a plane [18], and a sphere-on-cone model [19]. Usually only the apex field enhancement factor is calculated when dealing with a hemi-ellipsoid on a plane [18]. Influence of the anode-cathode distance within a floating sphere model was investigated in [20]. The model of a hemisphere on a post for CNT emitters was used in numerically simulations and reported in many papers [21-24]. Calculation difficulties in these numerical methods arise due to the large nanotube aspect ratio and very long distance between cathode and anode in comparison with emitter height. Usually, these numerical results were generalized and simple fitting formulas of field enhancement factor for individual nanotube [21, 22, 25, 26], for nanotube in space between parallel cathode and anode planes [27-30], and for a nanotube surrounded by neighboring nanotubes with a screening effect [31-35] were suggested.

The main problem for such algebraic fitting formulas is the lack of a definitive proof of their accuracy. Only a few papers considered forces acting on nanoemitters under electric field [32, 36-38]. Thus far, there is no analytical formula which provides a good approximation to the total current generated by the nanoscale field emitter. In this paper, we consider the electric field strength, field enhancement factor, ponderomotive forces, and total current of a metallic elliptical needle in the form of hemi-ellipsoid in the presence of a flat anode. We propose an exact analytical method, exploiting a very simple idea. The basic idea is to replace the cathode by a linear charge distribution in a uniform electric field and to use a set of ``image'' charges to reproduce the anode.
\section{Basic idea of analytical method}
Consider a prolate metallic spheroid in a uniform electric field. We can replace the spheroid by a linearly charged thread. The thread is a green line on Figure 1 and the linear charge distribution is represented by a red line. The length of a thread is $2h$.
\begin{figure}
\includegraphics[width=150pt]{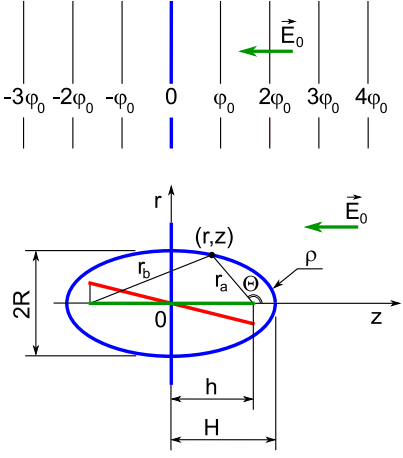}
\caption{Linearly charged thread in a uniform electric field along $z$.}
\end{figure}
The electrostatic potential produced by the charged thread is:
\begin{equation}
 \varphi(r,z)=-\int\limits_{-h}^h \frac{1}{4\pi\varepsilon_0} \frac{\tau z'\,dz'}{\sqrt{\bigl(z'-z\bigr)^2+r^2}},
\end{equation}
where $\varepsilon_0$ is the electric permittivity, ($r,z$) denotes the in-plane radial and $z$ coordinates, $\tau z$
is the linear charge density at point ($0,z$), $h$ is half of the thread length. The solution is independent of the azimuthal angle. We assume the coefficient of linear charge density, $\tau$ to be positive.

After integration we have
\begin{multline}
\varphi(r,z)=\frac{\tau z}{4\pi\varepsilon_0}
\biggl[\frac{4h}{\sqrt{(z-h)^2+r^2}+\sqrt{(z+h)^2+r^2}}-\\
\ln\biggl(
\frac{\sqrt{(z-h)^2+r^2}+h-z}{\sqrt{(z+h)^2+r^2}-h-z}
\biggr)\biggr].
\end{multline}
The total potential is
\begin{equation}
\Phi(r,z)=\varphi(r,z)+E_0 z
\end{equation}
where $E_0$ is the applied macroscopic field or far field. The shape of the metallic spheroid is given by the solution to the equation
\begin{equation}
\varphi(r,z)+E_0 z=0.
\end{equation}
Using new variables $r_a=\sqrt{(z-h)^2+r^2}$ and $r_b=\sqrt{(z+h)^2+r^2}$ and the equality $z=\frac{r_b^2-r_a^2}{4h}$,
we can rewrite Eq.~(4) in the form
\begin{equation}
\frac{4h}{r_a+r_b}-\ln\biggl(\frac{r_a+r_b+2h}{r_a+r_b-2h}\biggr)=-C,
\end{equation}
where $C=4\pi\varepsilon_0\frac{E_0}{\tau}$.
From this equation we obtain $r_a + r_b =\mbox{const}$, which means the cross-section of the spheroid is truly an ellipse. Points ($0,-h$) and ($0,h$) are the focii of the ellipse. Therefore $r_a$ and $r_b$ are distances between ($r,z$) and the two focii.
\section{Geometry of the ellipsoidal tip of the metallic needle}
Using a dimensionless parameter, the eccentricity
\begin{equation}
\xi=\frac{2h}{r_a+r_b}\:(0<\xi<1)
\end{equation}
we transform Eq.~(5) into
\begin{equation}
C=\ln\biggl(\frac{1+\xi}{1-\xi}\biggr)-2\xi.
\end{equation}
If $\xi$ is close to 1, the ellipse becomes elongated. If $\xi\rightarrow 0$ the ellipse turns into a circle. Therefore, by changing the coefficient of linear charge density, $\tau$  we may modify the shape of the ellipse.

We can also adjust other geometrical parameters of the ellipse: the length of semi-major axis or height $H$; the length of semi-minor axis or base radius, $R$ at $z=0$; and radius of curvature, $\rho$ at point ($0,H$) (see Fig.~1).
\begin{equation}
H=\frac{h}{\xi}=\frac{r_a+r_b}{2},\:R=\frac{h\sqrt{1-\xi^2}}{\xi},\:\rho=\frac{R^2}{H}.
\end{equation}
\section{Electric field and field enhancement factor on the surface of the metallic spheroid}
Taking the gradient of Eq.~(3) and using Eq.~(4) we can calculate components of the electric field on the surface of the metallic spheroid:
\begin{eqnarray}
E_z=-\frac{\partial\Phi(r,z)}{\partial z}=-\frac{E_0}{C}\frac{h(r_b-r_a)^2}{r_ar_b(r_a+r_b)},\\
E_r=-\frac{\partial\Phi(r,z)}{\partial r}=\phantom{mmmmmmmmmm} \nonumber\\
\hfil-\frac{E_0}{C}\frac{h(r_b-r_a)\sqrt{4h^2-(r_b-r_a)^2}}{r_ar_b\sqrt{(r_a+r_b)^2-4h^2}}.
\end{eqnarray}
Thus the modulus of the electric field is
\begin{equation}
E=\frac{E_0}{C}\frac{4h^2(r_b-r_a)}{(r_a+r_b)\sqrt{r_ar_b\bigl[(r_a+r_b)^2-4h^2\bigr]}}.
\end{equation}
Eq.~(9-11) allow determining the electric field strength on the surface of the half ellipsoid at an arbitrary point. The field enhancement factor at the apex of the ellipsoid follows from Eq.~(11)
\begin{equation}
\beta=\frac{2\xi^3}{(1-\xi^2)C}=\frac{2\xi^3}{(1-\xi^2)\left(\ln\frac{1+\xi}{1-\xi}-2\xi\right)}.
\end{equation}

Analytical expressions for field strength on the z axis and for field enhancement factor on the tip of the half ellipsoid obtained previously [18, 39-41] are in agreement with our result. Here, by taking gradient of Eq.~(3) we can obtain the field strength at any point we desire. In the limit $\xi\rightarrow 0$ we have a metallic half sphere and the field enhancement factor $\beta=3$.
If $\xi\rightarrow 1$ then for the elongated metallic needle, we have
\begin{equation}
\beta=2\frac{H}{\rho}\left[\ln\biggl(\frac{4H}{\rho}\biggr)-2\right]^{-1}.
\end{equation}
\begin{figure}
\includegraphics[width=200pt]{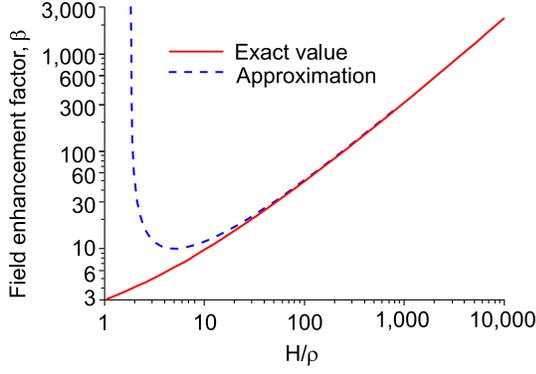}
\caption{Comparison of the exact formula (12) with approximate formula (13) for the field enhancement factor versus $H/\rho$.}
\end{figure}
In Fig.~2 we see that the approximate Eq.~(13) has a vertical asymptotic line at $H/\rho=e^2/4\approx 1.85$. The exact Eq.~(12) gives 3 for the half sphere when $H/\rho\rightarrow 1$. However, when $H/\rho>100$, the relative error becomes less than 1.7\%.
\section{Ponderomotive forces}
Let us estimate the electric force acting on the surface of ellipsoid. The electrostatic force acting on the elementary area, s of the external surface is given by
\begin{equation}
\vec{F}=\int\limits_S\frac{\varepsilon_0}{2}E^2\vec{n}\,ds,
\end{equation}
where $\vec{n}$ is a vector normal to the surface.

We can calculate the force acting on the spheroid surface between circles $r_a=A$ and $r_a=B$ (see Fig.~3).
\begin{figure}
\includegraphics[width=120pt]{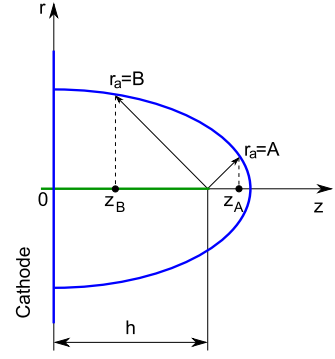}
\caption{Geometry for the calculation of ponderomotive force acting on the belt between $r_a=A$ and $r_a=B$.}
\end{figure}
It is obvious that $r$-component of force on that surface is equal to zero. For the $z$-component, after routine operations we obtain
\begin{equation}
dF_z=\frac{\varepsilon_0}{2}E^2n_z\,ds=\frac{\pi\varepsilon_0}{2}E^2\,dr^2=\frac{\pi\varepsilon_0}{2}E^2\frac{dr^2}{dr_a}\,dr_a.
\end{equation}
We can integrate the above equation analytically for $r_a$ changing from $A$ to $B$. We have
\begin{multline}
F_z(A|B)=\frac{\pi\rho^2\varepsilon_0(\beta E_0)^2}{2\left(1-\frac{\rho}{H}\right)^2}\times\\
\biggl[\ln\biggl(\frac{2r_a}{H}-\frac{r_a^2}{H^2}\biggr)-\frac{2r_a}{H}+\frac{r_a^2}{H^2}\biggr]\biggm|_A^B.
\end{multline}
Using $r_a=H-z\xi$ we rewrite
\begin{multline}
F_z(z_A,z_B)=\frac{\pi\rho^2\varepsilon_0(\beta E_0)^2}{2(1-\rho/H)^2}\times\\
\biggl[
1-\biggl(1-\frac{\rho}{H}\biggr)\frac{z^2}{H^2}-
\ln\biggl\{1-\biggl(1-\frac{\rho}{H}\biggr)\frac{z^2}{H^2}\biggr\}
\biggr]\biggm|_{z_A}^{z_B}.
\end{multline}
where $z$ varies from $z_A$ to $z_B$ (see Fig. 3). It gives us the value of net detaching force acting on the surface of the spheroid integrated over the surface between planes $z=z_A$ and $z=z_B$. The total detaching force acting on the ellipsoidal needle is
\begin{equation}
F_{total}=\frac{\pi\rho^2\varepsilon_0(\beta E_0)^2}{2(1-\rho/H)^2}\biggl(\frac{\rho}{H}-1-\ln\frac{\rho}{H}\biggr).
\end{equation}
\begin{figure*}
\includegraphics[width=350pt]{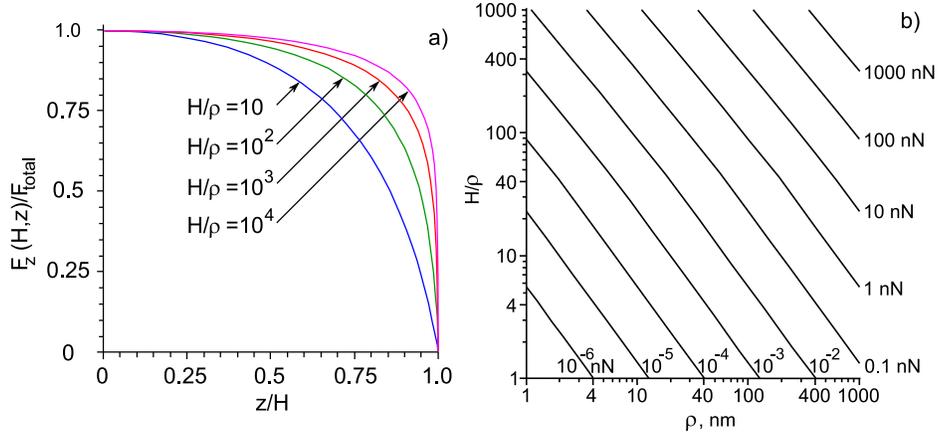}
\caption{Distribution of force and total force isolines. {\bf(a)} Distribution of relative force over the axis of needle. {\bf(b)} Isolines of total force on the  plane.}
\end{figure*}

In Fig. 4a we show the relative detaching force $F_z(H,z)/F_{total}$ as a function of coordinate $z$, where $F_z(H,z')$ is the force acting on the surface between plane $z=z'$ and the tip of the spheroid ($z=H$). Distribution depends only on the parameter $H/\rho$. The major part of the detaching force is concentrated near the tip when $H/\rho$ is large. Fig. 4b shows the total force isolines on the ($\rho,H/\rho$) plane. When we chose logarithmic scale for $\rho$ and $H/\rho$ with logarithmic steps we obtain a set of nearly straight isolines with equal distances in the plot.
\section{Field emission from individual needle}
\begin{figure*}
\includegraphics[width=350pt]{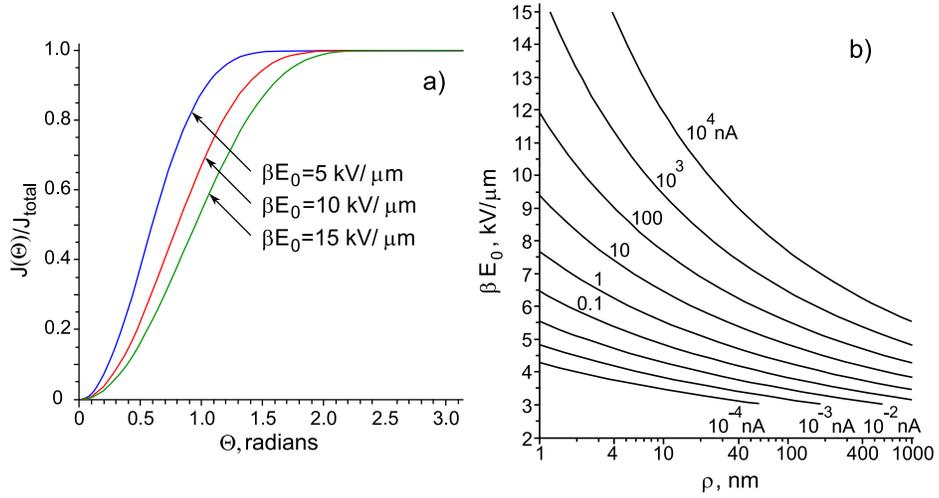}
\caption{Distribution of emission current and total emission current isolines for work function $\phi=4.8\:\mbox{eV}$ {\bf(a)} Distribution of relative emission current over $\Theta$. {\bf(b)} Total emission current isolines on the ($\beta E_0,\rho$) plane.}
\end{figure*}
The field emission from a metallic surface is described by the Fowler-Nordheim theory. According to this theory
\begin{equation}
j\approx\frac{C_1E^2}{\phi}\exp\biggl(-\frac{C_2\phi^{3/2}}{E}\biggr),
\end{equation}
where $j$ denotes the emission current density in $\mbox{A}\cdot\mbox{cm}^{-2}$, $E$ is local electric field at the emitting surface in $\mbox{V}\cdot\mbox{cm}^{-1}$, $\phi$ is work function in eV, and the first and second Fowler–Nordheim constants are $C_1=1.56\times10^{-6}[\mbox{A}\cdot\mbox{eV}\cdot\mbox{V}^{-2}]$, $C_2=6.83\times10^7[\mbox{V}\cdot\mbox{eV}^{-3/2}\cdot\mbox{cm}^{-1}]$ respectively.

To integrate Eq.~(21) over the surface, we express the elementary area as the following $ds=2\pi y\,dl=2\pi r_a\sin\theta r_a\,d\theta=-2\pi r_a^2\,d\lambda$, where $\lambda=\cos\theta$, $\theta$ is angle between axis $z$ and $r_a$ as shown in Fig. 1. Ellipse equation in polar coordinates is
\begin{equation}
r_a=\frac{\rho}{1+\xi\lambda},
\end{equation}
where $\lambda\in(-\xi,1)$ gives half of the ellipse which we are interested.
In the case of half sphere we have
\begin{equation}
\beta=3,\:E=3E_0\lambda,\:\rho=H.
\end{equation}
The current emitted from the surface bounded by $\theta\in(0,\Theta)$ is
\begin{widetext}
\begin{equation}
J(\Theta)=2\pi\rho^2\frac{9C_1E_0^2}{\phi}\int\limits_{\cos\Theta}^1 \lambda^2\exp
\biggl(-\frac{C_2\phi^{3/2}}{3E_0\lambda}\biggr)\,d\lambda=\\
\frac{2\pi\rho^2C_1\phi^{7/2}C_2^3}{3E_0}\biggl[
\biggl(\frac{\lambda^3}{3C_3^3}-\frac{\lambda^2}{6C_3^2}+\frac{\lambda}{6C_3}\biggr)
\exp\biggl(-\frac{C_3}{\lambda}\biggr)-\\
E_1\biggl(\frac{C_3}{\lambda}\biggr)\biggr]
\biggm|_{\lambda=\cos\Theta}^1,
\end{equation}
\end{widetext}
where $C_3=\frac{C_2\phi^{3/2}}{3E_0}$ and $\mbox{E}_1(x)\equiv\int\limits\frac{\exp(-xt)}{t}\,dt$ is the exponential integral. Due to small $\beta$ for the half sphere we need to use very strong electrical field to produce slightly visible current in experiment. In the general case, the exact formula for emission current should be written as
\begin{multline}
J(\Theta)=\frac{2\pi C_1}{\phi}\frac{(\beta E_0)^2\rho}{H-\rho}\int_{\cos\Theta}^1
\frac{r_a(H-r_a)^2}{2H-r_a}\times\\
\exp\biggl(-\frac{C_2\phi^{3/2}\sqrt{r_a(H-\rho)(2H-r_a)}}{\beta E_0(H-r_a)\sqrt{\rho}}\biggr)\,d\lambda,
\end{multline}
where for $r_a$ we use relation (20). Unfortunately it is not possible to integrate (23) in the general case, but we can obtain approximate formula for the elongated ellipsoidal needle. In this case the emission takes place mostly from the tip of the needle where the field is maximal due to the curvature being the largest there.

Due to $\rho\ll H$ we have
\begin{equation}
\xi\approx 1\Rightarrow r_a\approx\frac{\rho}{1+\lambda}.
\end{equation}
For the evaluation of (23) we assume $H-r_a\approx H$, $2H-r_a\approx 2H$. Then the integral can be expressed in terms of the $\mbox{E}_1$-function and we get
\begin{equation}
J(\Theta)=\frac{2\pi\rho^2C_1(\beta E_0)^2}{\phi}\mbox{E}_1\biggl[\frac{C_2\sqrt{2}\phi^{3/2}}{\beta E_0\sqrt{1+\lambda}}\biggr]\biggm|_{\lambda=\cos\Theta}^1.
\end{equation}
By comparing with accurate numerical integration of (23) we find that formula (25) is accurate up to the first four digits for $H/\rho>100$. So formula (25) is a good approximation for investigating the current emitted from an area of surface depending on angle $\Theta$. The total current with similar accuracy can be written approximately as
\begin{multline}
J_{total}=\frac{2\pi\rho^2C_1(\beta E_0)^2}{\phi} \mbox{E}_1
\biggl[\frac{C_2\sqrt{2}\phi^{3/2}}{\beta E_0 \sqrt{1+\lambda}}\biggr]\biggm|_{-\xi}^1\approx\\
\frac{2\pi\rho^2C_1(\beta E_0)^2}{\phi}\mbox{E}_1\biggl[\frac{C_2\phi^{3/2}}{\beta E_0}\biggr].
\end{multline}
\begin{figure*}
\includegraphics[width=350pt]{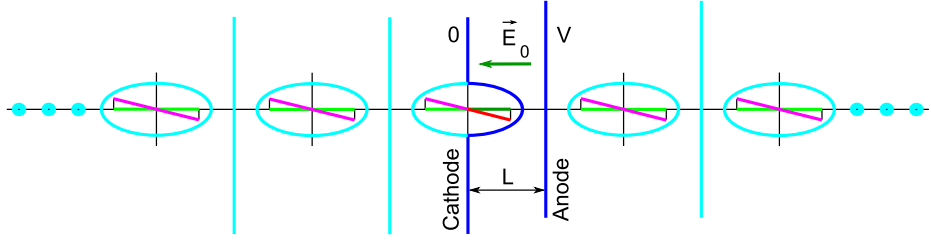}
\caption{Infinite set of ``image'' charges for the simulation of a planar anode.}
\end{figure*}
The emission current has much sharper distribution on the tip than the detaching force due to exponential dependence of Fowler-Nordheim formula (19). In Fig. 5a we plot the relative emission current, $J(\Theta)/J_{total}$ as a function of $\Theta$ (angle between $r_a$ and axis $z$), where $J(\Theta)$ is emission from the tip on the surface of the ellipsoidal needle. From (26) we see that total emission current depends only on the area of the hemisphere, $2\pi\rho^2$, work function, $\phi$, and the electric field on the tip, $\beta E_0$. So in Fig. 5b we show the total emission current isolines within a reasonable range (from $10^{-4}$ to $10^4$ nA) with logarithmic steps on the ($\beta E_0,\rho$) plane, where the work function $\phi=4.8\mbox{ eV}$ is assumed. If the electric field is $E_0\approx 1..5\:\mbox{V}/\mu\mbox{m}$, then we have to use an enhancement factor $\beta=10^3..10^4$ to get an appreciable current.

\section{Simulation of planar anode by ``image'' charges}

The presence of a flat anode placed at a distances comparable with the length of the nano needle has strong influence on the emission current and the detaching force. The basic idea of calculation is to replace the cathode by a linearly charged thread in a uniform electric field and to use a set of ``image'' charges to reproduce the anode as shown in Fig. 6. We put infinite set of ``images'' of the linearly charged thread with the same spacing, $2L$ ($L>H$). It is clear that planes $z=0$ and $z=L$ are planes of symmetry for distributed charges and will have potentials 0 on cathode and $V=E_0L$ on anode.
With the same $\tau$ we will get different geometry of cathode due to additional potential of ``images''. We assume that the new form of the elongated needle will be approximately described by ellipsoid especially near the tip. On the surface of the thin ellipsoid, we have $r\ll h$ and $z<H<L$. So, we can describe the image potential $\varphi_i$ as a function of $z$.
\begin{multline}
\varphi_i(z)=-\frac{E_0}{C^*}\sum\limits_{n=1}^\infty\biggl[
(z-2nL)\ln\frac{2nL-z+h}{2nL-z-h}+\\
(z+2nL)\ln\frac{2nL+z+h}{2nL+z-h}\biggr],
\end{multline}
where $C^*\equiv 4\pi\varepsilon_0 E_0/\tau^*$, $L=\eta H$, $\eta>1$. The thread potential is
\begin{equation}
\varphi=\frac{zE_0}{C}\biggl[\frac{4h}{r_a+r_b}-\ln\biggl(\frac{r_a+r_b+2h}{r_a+r_b-2h}\biggr)\biggr].
\end{equation}
Near the tip $z\approx H$ we have the following equation for describing the shape of the ellipsoid:
\begin{equation}
\varphi(r,z)+E_0z+\varphi_i(z)\approx\varphi(r,z)+E_0z+\frac{z}{H}\varphi_i(H)=0.
\end{equation}
Then in analogy to (5) we write
\begin{equation}
\frac{4h}{r_a+r_b}-\ln\biggl(\frac{r_a+r_b+2h}{r_a+r_b-2h}\biggr)=-C^*-P,
\end{equation}
where
\begin{multline}
P=\sum_{n=1}^\infty\biggl[(2n\eta-1)\ln\frac{2n\eta-1+\xi}{2n\eta-1-\xi}-\\
(2n\eta+1)\ln\frac{2n\eta+1+\xi}{2n\eta+1-\xi}\biggr].
\end{multline}
So we have
\begin{equation}
C^*=\ln\biggl(\frac{1+\xi}{1-\xi}\biggr)-2\xi-P.
\end{equation}
Analogous to (9) and (10) we have
\begin{multline}
E_z=-\frac{\partial\Phi(r,z)}{\partial z}\approx-E_0-\frac{\partial\varphi}{\partial z}
-\frac{\partial\varphi_i}{\partial z}\biggm|_{z=H}=\\
\frac{E_0}{C^*}\biggl(-\frac{h(r_a-r_b)^2}{r_ar_b(r_a+r_b)}+P-Q\biggr),
\end{multline}
\begin{equation}
E_r=-\frac{\partial\Phi(r,z)}{\partial r}=-\frac{E_0}{C^*}
\frac{h(r_b-r_a)\sqrt{4h^2-(r_b-r_a)^2}}{r_ar_b\sqrt{(r_a+r_b)^2-4h^2}},
\end{equation}
where
\begin{multline}
Q\equiv\sum_{n=1}^\infty\biggl[
\frac{8n\eta\xi(4n^2\eta^2-1-\xi^2)}{\bigl(4n^2\eta^2-(1+\xi)^2\bigr)
\cdot\bigl(4n^2\eta^2-(1-\xi)^2\bigr)}-\\
\ln\frac{(2n\eta+\xi)^2-1}{(2n\eta-\xi)^2-1}\biggr].
\end{multline}
If $|(P-Q)/(\beta\cdot C^*)|<0.01$ then we may neglect $(P-Q)/C^*\approx 0$ in (33). We get
\begin{equation}
\beta(L/H,H/\rho)=\frac{2\xi^3}{(1-\xi^2)C^*}.
\end{equation}

Finally for the electrical field with the assumption $(P-Q)/C^*\approx 0$ formulas (33) and (34) are completely analogous to (9) and (10). Therefore for the force and emission current we may use the above derived formulas. Instead of (12) we should use (31), (32), and (36). So, the parameter $\eta=L/H$ in the force and emission current formulas (17), (18), (25), and (26) should be included only through the field enhancement factor $\beta=\beta(L/H,H/\rho)$.
\begin{figure}
\includegraphics[width=130pt]{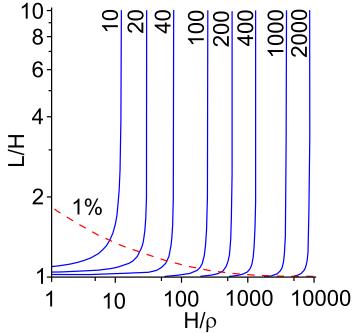}
\caption{Isolines for the field enhancement factor $\beta$ on the ($L/H,H/\rho$) plane.}
\end{figure}
\begin{figure}
\includegraphics[width=200pt]{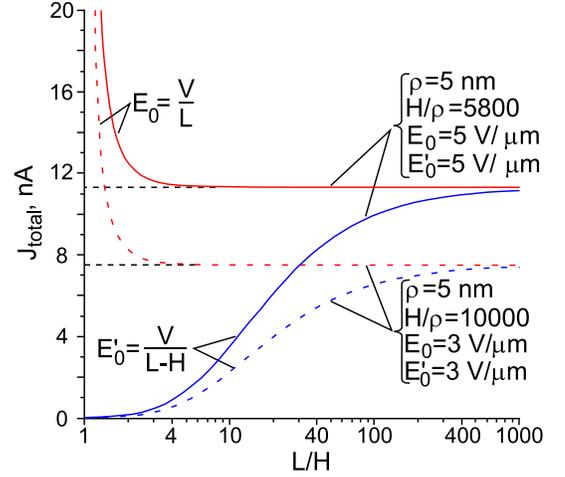}
\caption{Emission current as a function of the anode-cathode distance parameter $\eta=L/H$ for constant electric field $E_0\equiv V/L$ (red line) or $E'_0\equiv V/(L-H)$ (blue line). Radius of curvature is fixed at $\rho=5\:$nm. For solid curves, $H/\rho=5800$ and for dash curves, $H/\rho=10000$. For red and blue solid curves $E_0=5\mbox{V}/\mu\mbox{m}$, for red and blue dash curves $E_0=3\mbox{V}/\mu\mbox{m}$.}
\end{figure}
In Fig. 7 we present isolines for the field enhancement factor $\beta$ on the ($L/H,H/\rho$) plane. The dashed line divides the graph into the upper-right and lower-left side. On the upper-right side our approximation has less than 1\% error, $|(P-Q)/(\beta\cdot C^*)|<0.01$.

What distance between the anode and cathode is large enough to assert that the elongated metallic spheroid is placed in a uniform electric field? In experiment we may measure distance $L$ between the anode and cathode and the distance, $L-H$ between the anode and the needle apex. We can determine the applied electric field both by $E_0=V/L$ and by $E'_0=V/(L-H)$. In our calculations we move the anode and increase the voltage $V$ so as to keep $E_0$ or $E'_0$ immutable. It is clear that for a large distance (in the limit $L\rightarrow\infty$) the difference between definitions of $E_0$ and $E'_0$ disappears. Fig. 8 shows the total current emitted from the needle versus the anode-cathode distance.

We set for blue lines $E'_0=\mbox{const}$ and for red lines $E_0=\mbox{const}$. With $E'_0=E_0=\mbox{const}$ the total currents have the same limit for $\eta=L/H\rightarrow\infty$. But currents with $E'_0=\mbox{const}$ tend to approach the limit much slower than the corresponding currents with $E_0=\mbox{const}$. If we define the applied electric field as $E_0$ and the anode-cathode distance is ten times more than the needle height, then we can neglect the influence of the anode and assume that the metallic needle is placed in a uniform field. In contrast, if the applied electric field is defined as $E'_0$, then the anode-cathode distance must exceed one thousand times of the needle height for the above statement to be valid.

\section{Conclusion}

In this work we theoretically investigate the field emission from an elongated prolate spheroid metallic needle in diode configuration between a flat anode and cathode. Exact analytical formulas of the electrical field, field enhancement factor, ponderomotive force, and field emission current are found. Applied voltage, height of the needle, radius of curvature on its top, and the work function are the parameters at our disposal. The field enhancement factor, total force and emission current, as well as their distributions on the top of the needle for a wide range of parameters, have been calculated and analyzed.

\nocite{*}
\bibliography{Ellipsoidal_needle}
\end{document}